\documentclass[epj]{svjour}
\usepackage{ulem}
\usepackage{amssymb}
\usepackage{graphicx}
\usepackage{epsfig}
\usepackage{subfigure}
\usepackage{verbatim}
\usepackage{color}
\usepackage{enumerate}
\usepackage{bm}

\begin{document}

\title{On Solving Cubic-Quartic Nonlinear Schr\"odinger Equation in a Cnoidal Trap}

\author{Argha Debnath, Ayan Khan\thanks{ayan.khan@bennett.edu.in}}
   \institute{
   Department of Physics, School of Engineering and Applied Sciences, Bennett University, Greater Noida, UP-201310, India
   }


\abstract{
The recent observations of quantum droplet in ultra-cold atomic gases have opened up new avenues of fundamental research. The competition between mean-field and beyond mean-field interactions, in ultra-cold dilute alkali gases, are believed to be instrumental in stabilizing the droplets.
These new understanding has motivated us to investigate the analytical solutions of a trapped cubic-quartic nonlinear Schr\"odinger equation (CQNLSE). The quartic contribution in the NLSE is derived from the beyond mean-field formalism of 
Bose-Einstein condensate (BEC). To the best of our knowledge, a comprehensive analytical description of 
CQNLSE is non-existent. Here, we study the existence of the analytical solutions which are of the cnoidal type for CQNLSE. The external trapping plays a significant role in the stabilization of the system. 
In the limiting case, the cnoidal wave solutions lead to the localized solution of  bright solution and delocalized kink-antikink pair. The nonexistence of the sinusoidal mode in the current scheme is also revealed in our analysis.
}

\maketitle

\section{Introduction}
The dynamics of Bose-Einstein condensate (BEC) is quite successfully captured via mean-field formalism developed by Gross and Pitaevskii and therefore the nonlinear Schr\"odinger equation (NLSE) associated with BEC dynamics is known as Gross-Pitaevskii (GP) equation \cite{gross,pitae}. Since the first successful realization of the atomic BEC, numerous works in this unique quantum state have taken place \cite{stringari1999,stringari2008,bloch} which has widened our understanding about this intriguing phase of matter.
 
Very recently, a group of experimentalists has observed the formation of liquid droplet like state in a BEC mixture \cite{cabrera1}. This is quite surprising as the prevailing conception of the liquid state is highly influenced by the theory of van der Waals. It asseverates that the liquid state arises when density is quite high due to an equilibrium between attractive inter-atomic forces and short-range repulsion. However, these newly emerged droplets in ultra-cold and extremely dilute atomic gases do not explicitly follow the usual van der Waals predicted perception  \cite{barbut2}. This liquid like state is purely a manifestation of the quantum fluctuations \cite{barbut1,barbut3}.
These droplets are small clusters of atoms, self-bound by the interplay of attractive and repulsive forces. The origin of the attractive force can be modeled in the purview of standard mean-field theory whereas the repulsive force originates from the beyond mean-field correction \cite{sala1,arlt}. The underlying theory relies on the Lee-Huang-Yang's (LHY) correction \cite{lee} to the mean-field GP equation. 

From the mathematical perspective, the problem boils down to a nonlinear equation whose nonlinearity is not limited to the cubic term but it also carries a quartic contribution in three dimensions \cite{cabrera2}. From our survey of literature we realize that there exists a significant void in understanding the competition between cubic and quartic nonlinearity. We were only able to find out some discussion in Ref.~\cite{inui,yamano}, where the authors have studied the existence of soliton in cubic-quartic Nonlienar Schr\"odinger equation (CQNLSE) by using phase portrait analysis.

Nonlinear Schr\"odinger equation has drawn interest from diverse communities for the last several decades ranging from water waves to plasma. Formally, NLSE is the homogeneous second-order nonlinear differential equation and they admit different classes of the analytical solution. It is not difficult to map the NLSE to Jacobi Elliptic equation \cite{atre} which in turn allows us to use the 12 Jacobi elliptic functions as solutions. It must be noted that, the Jacobi elliptic functions can be derived from the amplitude function of Jacobi elliptic integrals \cite{abro,takeuchi}.
These solutions can be constant, periodic, or localized based on the parameter $m$ as $0\leq m\leq1$. A broad class of the localized solutions are categorized as the solitons which are highly sought after in fiber optic communication system, as robust localized pulses with the ability to retain the shape over a large distance is highly amenable for long-distance communication \cite{govind}.
Apart from this, one can also find solitons in bosonic and fermionic superfluids \cite{streker,khyko,spun,khan4,yefsah,ren} where these localized  matter waves can play an useful role in different fields such as atom lithography and atom optics.

In this article, our primary interest is to obtain analytical solutions for CQNLSE in a cigar-shaped BEC. The systematic dimensional reduction mechanism can be employed to reduce the 3+1 dimensional problem to a 1+1 dimensional problem. Using the prescription of Ref.\cite{atre} we can actually reduce 3+1 dimensional CQNLSE described in Ref.\cite{cabrera2} to a quasi one-dimensional CQNLSE. Here, we will start from a trapped quasi one dimensional CQNLSE. We realize that the external trap plays a pivotal role in stabilizing the solution in quasi one-dimensional geometry.

Since regular NLSE posses cnoidal solutions \cite{cervero}, therefore we first attempt to obtain a static cnoidal solution in CQNLSE which is trapped in a cnoidal potential. Later we will explicate the physical nature of the potential. 
In Sec.~\ref{model} we derive the cnoidal solutions associated with a trapped CQNLSE. Here, we like to point out that, to the best of our knowledge these solutions are provided for the very first time. Next, we concentrate on the stability of the obtained solutions in Sec.~\ref{stability} and extend our analysis to traveling waves and show that our static analysis can be mapped in the dynamical system as well with suitable Galilean transformation in the moving frame. We elaborate on this description in Sec.~\ref{dyn}. We draw our conclusion in Sec.~\ref{con}.

\section{Introduction to CQNLSE}\label{model}
Based on our discussion in the previous section, we realize that, understanding the interplay of cubic and quartic nonlinearity in a CQNLSE is the need of the hour. Though the recent experiments were performed for binary BEC \cite{cabrera1} and dipolar BEC \cite{barbut3}, we consider a CQNLSE in the presence of an external trap in a quasi one-dimensional geometry. Since, there exists a substantial void in the analytical description of the CQNLSE, we choose to start from the relatively simple one-dimensional geometry and provide necessary analytical insight. In this paper, our main is to obtain cnoidal solutions from the CQNLSE therefore we set the external potential trap such a way that it allows the cnoidal solution to stabilize. 
Our starting point can be noted as, 
\begin{eqnarray}
i\frac{\partial\psi(x,t)}{\partial t}&=&-\mathcal{A}\frac{\partial^2\psi(x,t)}{\partial x^2}+\mathcal{B}|\psi(x,t)|^2\psi(x,t)+\nonumber\\
&&\mathcal{C}|\psi(x,t)|^3\psi(x,t)+\mathcal{D}(x)\psi(x,t)\label{eq1}
\end{eqnarray}
Here, $\mathcal{A}$, $\mathcal{B}$ and $\mathcal{C}$ are coefficients. In GP equation $\mathcal{A}$ is equivalent to $\hbar^2/2m$, where as in nonlinear fiber optics it defines the dispersion. $\mathcal{B}$ and $\mathcal{C}$ are the strength of the nonlinearities, where $\mathcal{B}$ can be connected to the short range two-body $s$-wave scattering and $\mathcal{C}$ defines the beyond mean field correction.  $\mathcal{D}$ is the external potential. Here, we are interested in analyzing the static solutions and for that purpose we define, $\psi(x,t)=\phi(x)e^{-i\mu t}$.
So, taking into account the above considerations, Eq.(\ref{eq1}) leads to,
\begin{eqnarray}\label{eq2}
&&\frac{d^2\phi(x)}{dx^2}+\alpha\phi(x)-\beta|\phi(x)|^2\phi(x)-\gamma|\phi(x)|^3\phi(x)\nonumber\\&&-\delta(x)\phi(x)=0,
\end{eqnarray}
where, $\alpha=\mu/\mathcal{A}$, $\beta=\mathcal{B}/\mathcal{A}$, $\gamma=\mathcal{C}/\mathcal{A}$ and $\delta(x)=\mathcal{D}(x)/\mathcal{A}$. It must be noted that at this point, we are considering both the interactions (mean-field and beyond mean-field) 
as repulsive in nature. However, their exact characteristic can be understood after we obtain the exact solution. 

\section{Analytical Treatment of CQNLSE}
From our prior knowledge about NLSE we know that, it is possible to map NLSE to Jacobi elliptic equation \cite{atre,takeuchi} which poses a variety of solutions in the form of 12 Jacobian elliptic functions \cite{abro}. 
We assume an ansatz in the form of cnoidal solution as Eq.(\ref{eq2}) can be mapped to NLSE when the quartic nonlinearity is absent and no external force is applied. 
Since, cnoidal functions can lead to localized as well as sinusoidal modes based on the parameter value $m$, therefore, it is always beneficial if we can obtain a set of cnoidal solutions. We are primarily interested in finding the solutions in terms of the copolar trio, i.e.,  ``cn'', ``sn'' and ``dn''. Further, we know that the cnoidal functions at $m=0$ and $1$ can be written as, $\textrm{cn}(x,0)=\cos{x}$, $\textrm{cn}(x,1)=\textrm{sech}\,x$, $\textrm{sn}(x,0)=\sin{x}$, $\textrm{sn}(x,1)=\tanh{x}$, $\textrm{dn}(x,0)=1$ and $\textrm{dn}(x,1)=\textrm{sech}\,x$. Moreover, the derivatives are defined as, $\frac{d}{dx}\textrm{sn}(x,m)=\textrm{cn}(x,m)\textrm{dn}(x,m)$, $\frac{d}{dx}\textrm{cn}(x,m)=-\textrm{sn}(x,m)\textrm{dn}(x,m)$ and $\frac{d}{dx}\textrm{dn}(x,m)=-m\,\textrm{sn}(x,m)\textrm{cn}(x,m)$. 
Hence, a careful evaluation of the above relations reveal that $\textrm{dn}(x,1)=\textrm{cn}(x,1)$, $\textrm{dn}(x,0)$ is constant and derivative of $\textrm{dn}(x,0)$ is zero. 
This allows us to concentrate only on ``cn'' and ``sn'' solutions without any loss of generality. 

\subsection*{``cn'' solution}
Let us define the first ansatz of the form $\phi(z)=A+B\,\textrm{cn}(z,m)$ where $z=\zeta x$ and $\zeta$ is the inverse of coherence length. We consider the external potential as $\delta(z)=V_0\,\textrm{cn}^3(z,m)$, where $V_0$ is the strength of the external potential.
The external potential actually plays a crucial role in stabilizing the system against the competing nonlinearities. The potential can be induced by means of an optical laser field. At this point, we are not explicating the physical nature of the potential however, after obtaining the solution we will definitely do so.
Due to the change of variable, Eq.(\ref{eq2}) modifies to 
\begin{eqnarray}\label{eq3}
&&\zeta^2\frac{d^2\phi(z)}{dz^2}+\alpha\phi(z)-\beta|\phi(z)|^2\phi(z)-\gamma|\phi(z)|^3\phi(z)\nonumber\\&&-\delta(z)\phi(z)=0.
\end{eqnarray}

Applying the ansatz in Eq.(\ref{eq3}) we obtain a set consistency condition,
\begin{eqnarray}\label{coef_eq}
&&A^3\gamma+A^2\beta-\alpha=0\nonumber\\
&&4A^3\gamma+3A^2\beta-\alpha-(2m-1)\zeta^2=0\nonumber\\
&&4AB^3\gamma+B^3\beta+2m\zeta^2+AV_0=0,\nonumber\\
&&A=-\frac{\beta}{2\gamma},\,\, V_0=-B^3\gamma.
\end{eqnarray}
\begin{figure}\centering
\includegraphics[scale=0.25]{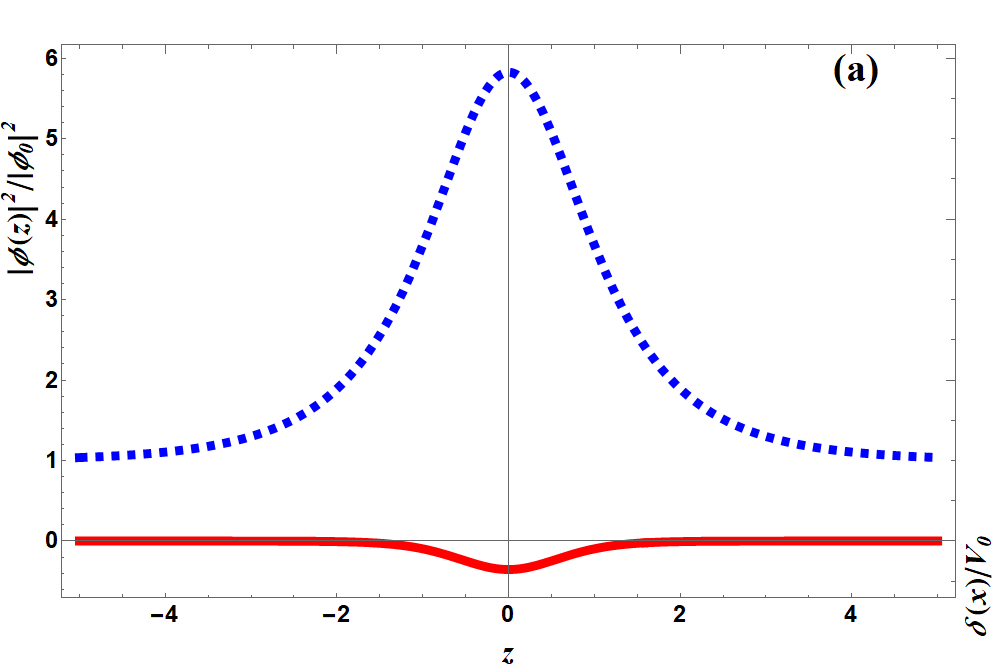}\\
\includegraphics[scale=0.25]{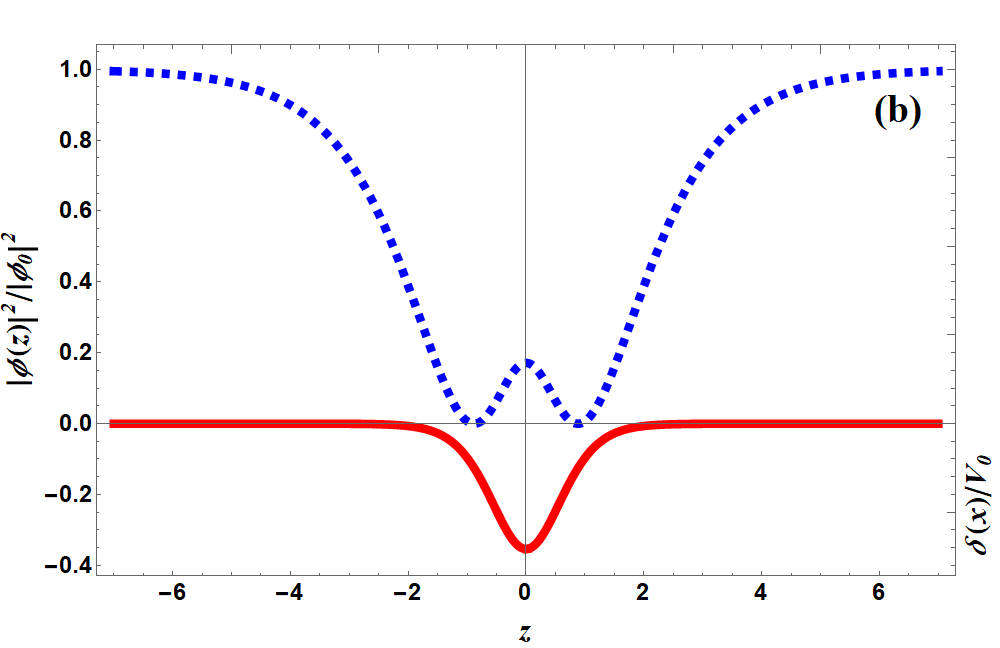}
\caption{(Color Online) The dashed blue line in both the figures depicts two localized solutions representing Eq.(\ref{sech}). The solid red line describes the potential profile which is similar to the P\"oschl-Teller potential.  In (a) we observe bright soliton like structure however, (b) describes exotic $w$-soliton. Here, $\phi_0=\beta/2\gamma$ and $V_0=-\frac{\beta^3}{\gamma^2}$}\label{sech_plot}
\end{figure}
Solving Eq.(\ref{coef_eq}) we obtain,
$\zeta=\frac{B\sqrt{\beta}}{2\sqrt{m}}$ and $B=\pm \frac{\beta}{\sqrt{4-\frac{2}{m}}\gamma}$. Thus the strength of optical potential can now be evaluated as, $V_0=\mp\frac{\beta^3}{2\sqrt{2}(\frac{-1+2m}{m})^{3/2}\gamma^2}$.
Hence, the solution is, 
\begin{eqnarray}\label{cn_sol}
\phi(z)=-\frac{\beta}{2\gamma}\left(1\mp\frac{\sqrt{2m}}{\sqrt{2m-1}}\textrm{cn}(z,m)\right).
\end{eqnarray}
However, the solution is acceptable if and only if, $\beta^3-8\alpha\gamma^2=0$. It must be noted that the mean-field interaction is repulsive otherwise, the coherence length will be complex which is totally undesirable. However, cnoidal solution does exist for repulsive as well as attractive beyond mean-field interaction. This counter intuitive situation is supported solely by the external potential. From Eq.(\ref{cn_sol}) we observe that the solution does not exist for $m=1/2$ and $m=0$ leads to constant solution. Hence, a sinusoidal solution or ``cos(x)'' /``sin(x)'' type of solution cannot be obtained in this framework. In summary, we can conclude that it is possible to obtain a localized solution corresponding to $m=1$ however, the sinusoidal mode corresponding to $m=0$ is absent.


\subsection*{Localized Solution}
As discussed in the previous section, the cnoidal ``cn'' solution indicates the existence of localized modes as the solution exists for $m=1$. Hence, from Eq.(\ref{cn_sol}) we can write the solution as,
\begin{eqnarray}\label{sech}
\phi(z)=-\frac{\beta}{2\gamma}\left(1\mp\sqrt{2}\textrm{sech}(z)\right).
\end{eqnarray}
The external potential necessary to support the solution turns out to be $\delta(z)=-\frac{\beta^3}{2\sqrt{2}\gamma^2}\textrm{sech}^3(z)$. In Fig.\ref{sech_plot} we describe both solutions along with the potential profile. It must be noted that the P\"oschl-Teller like potential (it must also be noted that, regular P\"oschl-Teller type potential is $\propto\textrm{sech}^2(z)$, whereas here the potential is $\propto \textrm{sech}^3(z)$) actually allows to stabilize the localized solution as it is evident from the figures. Using $B=- \frac{\beta}{\sqrt{2}\gamma}$ we obtain bright soliton-like profile. However, contrary to the common perception, here we observe a non zero background density of the bright soliton. As $z\rightarrow\pm\infty$, $|\phi(z)|^2\rightarrow\beta^2/4\gamma^2$. Since two-body short-range interaction ($\beta$) and the beyond mean-field contribution ($\gamma$) are related to the scaled chemical potential ($\alpha$) via $\beta^3=8\alpha\gamma^2$, therefore we can actually relate the background density in terms of the scaled chemical potential as noted in Ref.~\cite{petrov2,roy1}. Similarly, we also obtained $w$-soliton for $B= \frac{\beta}{\sqrt{2}\gamma}$ having a background density of $\beta^2/4\gamma^2$. The emergence of $w$-soliton is a manifestation of the competition between the interactions and its coupling with the trap. Similar observation has already been reported for strong coupling BEC \cite{roy2}. The normalization of the system is defined as $\int_{-\infty}^{\infty}|\phi(z)|^2dz=N$ which leads to $N=\frac{(1+\pi/\sqrt{2})\beta^2}{\gamma^2}$ and $-\frac{(\pi/\sqrt{2}-1)\beta^2}{\gamma^2}$ corresponding to Fig.~\ref{sech_plot}(a) and (b) respectively.
However, as particle number is always positive therefore the $w$-soliton solution is not physically acceptable.
\begin{figure}\centering
\includegraphics[scale=0.2]{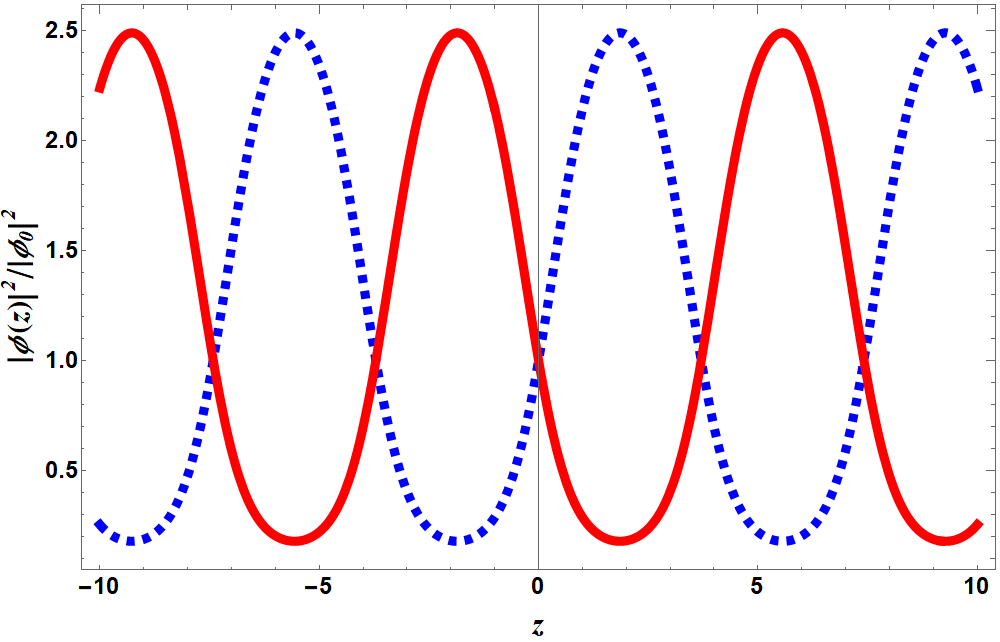}
\caption{(Color Online) The figure depicts periodic behavior of ``sn'' solution for $m=1/2$. The dashed blue line depicts $\phi_+(z)$ solution and the red solid line corresponds to the $\phi_-(z)$ solution from Eq.(\ref{sn_sol}). Here, $|\phi_0|^2=\beta^2/4\gamma^2$.}\label{sn_plot}
\end{figure}
\subsection*{``sn'' solution}
\begin{figure}\centering
\includegraphics[scale=0.2]{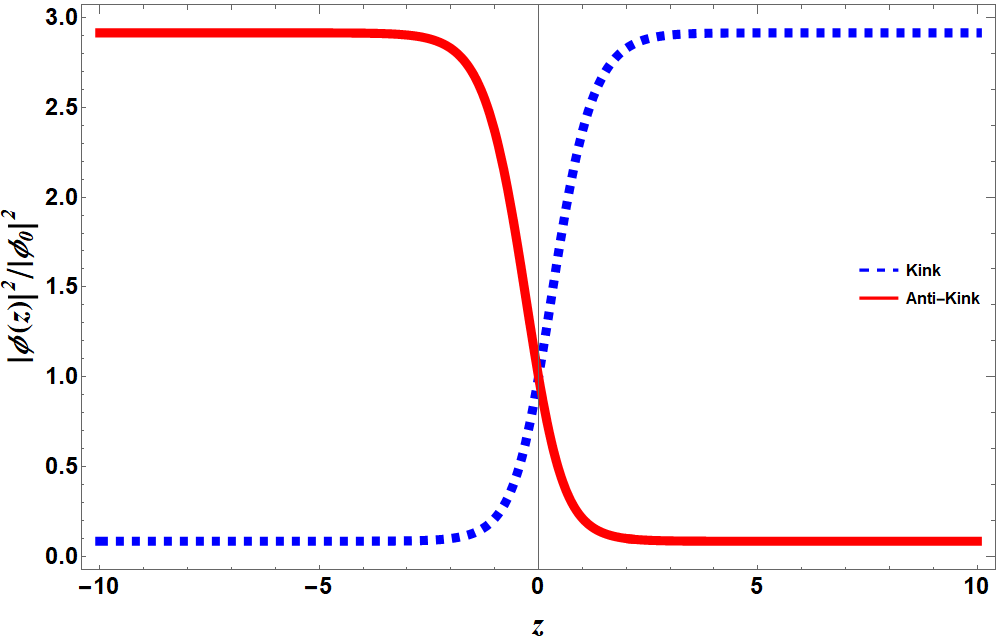}
\caption{(Color Online) The figure depicts kink and anti-kink solution obtained from Eq.(\ref{sn_sol}). Here, $|\phi_0|^2=\beta^2/4\gamma^2$.}\label{kink_plot}
\end{figure}

Next we use, $\phi(z)=A+B\,\textrm{sn}(z,m)$ as ansatz and $\delta(z)=V_0\,\textrm{sn}^3(z,m)$. Applying the ansatz and the cnoidal potential in Eq.(\ref{eq2}), we obtain, $A=-\beta/2\gamma$, $\alpha=\frac{\beta^3}{8\gamma^2}$, $\zeta=\frac{iB\sqrt{\beta}}{2\sqrt{m}}$, $B=\pm \frac{\beta\sqrt{m}}{2\sqrt{m+1}\gamma}$ and $V_0=\frac{\beta^3m^{3/2}}{8(m+1)^{3/2}\gamma^2}$. Here also we observe that for $m=0$ the coherence length reduces to zero as $\zeta\rightarrow\infty$ which implies nonexistence of sinusoidal modes. However, the two-body scattering length must be negative or attractive otherwise the coherence length will be an imaginary quantity. Hence, the cnoidal wave solution reads, 
\begin{eqnarray}\label{sn_sol}
\phi(z)&=&-\frac{\beta}{2\gamma}\left(1\mp\frac{\sqrt{m}}{\sqrt{m+1}}\textrm{sn}(z,m)\right).
\end{eqnarray} 
Contrary to the ``cn'' solution, we observe that ``sn'' solution yields nontrivial result for all values of $m$ except $m=0$ where the solution assimilates in the constant background. Fig.\ref{sn_plot} describes the behavior of the two solutions for $m=1/2$. The red solid line corresponds to $1-\frac{\sqrt{m}}{\sqrt{m+1}}\textrm{sn}(z,m)$ and the blue dashed line depicts $1+\frac{\sqrt{m}}{\sqrt{m+1}}\textrm{sn}(z,m)$. We note these two solutions as $\phi_-(z)$ and $\phi_+(z)$ respectively.
At $m=1$ limit we yield kink-anti kink solution as described in Fig.\ref{kink_plot}. Kink and anti-kink solitons are common in the context of Sine-Gordon (SG) equation \cite{dodd}. 
A common occurrence of Kink solution can be obtained in a ferromagnet describing the domain wall between two distinct spin domains. The magnetic spin rotates from one orientation in one domain to the opposite orientation in the adjacent domain. The transition region between the two opposite spins is called the Bloch wall.  Under the influence of an applied magnetic field, the Bloch wall can propagate according to the Sine-Gordon equation thereby yielding a Kink solution. Albeit, the current nonlinear system is not same as the SG equation, nevertheless the solution boils down to the kink and anti-kink pair.

\section{Stability}\label{stability}
In the previous section, we have elaborated the existence of different types of solitonic solutions. The main objective of this section is to investigate the stability of the obtained soliton solutions. Here, we plan to employ the well-known Vakhitov-Kolokolov (VK) criterion \cite{vakhitov1973stationary} for the localized solutions.  The VK criterion has been widely used in determining the stability of the solutions of the NLSE, which predicts the parameter regime in the chemical potential where the soliton’s amplitude can grow or decay exponentially \cite{roy1,roy2,das2}. The VK criterion states that a necessary stability condition is a positive slope in the dependence of the number of atoms on the chemical potential. If, $\mathcal{N_{\mu}} = \partial N/\partial \mu> 0$, the solution is found to be stable and for $\mathcal{N_{\mu}} < 0$, the solution is unstable. One must note that the condition $\mathcal{N_{\mu}}= 0$ provides the instability threshold, $\mu = \mu_{th}$ \cite{das2,pelinovsky1996,sakaguchi2010}.

It must be noted from Fig.~\ref{sech_plot} the asymptotic value of the solution leads to a flat bulk region contrary to the usual single soliton solution of NLSE \cite{petrov2}.  Hence the number of atoms in the condensate, after subtracting the suitable background, can be defined as, 
\begin{eqnarray}
N&=&\int_{-\infty}^{\infty}(\rho-\rho_0)dx\nonumber\\
&=&\frac{(1+\pi/\sqrt{2})\beta^2}{\gamma^2}\quad\textrm{for bright soliton}\label{ndb}\\
&=&-\frac{(\pi/\sqrt{2}-1)\beta^2}{\gamma^2}\quad\textrm{for $w$-soliton}.\label{nw}
\end{eqnarray}
Naturally, Eq.(\ref{nw}) is physically unacceptable hence we will concentrate on the stability of the bright soliton only.
As $\alpha$ is the scaled chemical potential we can calculate $\mathcal{N}_{\alpha}=\frac{\partial N}{\partial\alpha}=\frac{8(1+\pi/\sqrt{2})}{\beta}$.  Based on our previous discussion we know that $\beta$ is always positive, hence $\mathcal{N}_{\alpha}$
is also positive. This allows us to conclude that the bright soliton is stable under the linear stability analysis.
\begin{figure}\centering
\includegraphics[scale=0.2]{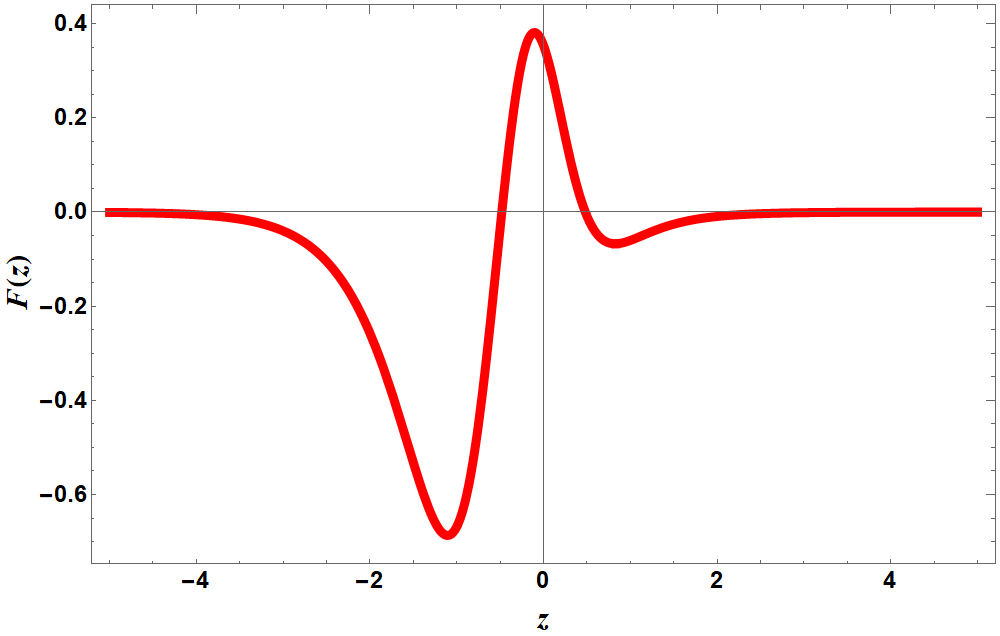}
\caption{(Color Online) Variation of effective force $F(z)$. Here, we have chosen an arbitrary value of $\beta=-1$ and magnitude of $\gamma$ is half the magnitude of $\beta$ without loss of any generality.}\label{force}
\end{figure}

To determine the stability criterion for kink soliton we need to analyze the effective potential $U(z)$ which can be computed from the fact $\zeta^2\frac{d^2\phi(z)}{dz^2}=-\frac{dU(z)}{d\phi}$. Here, $U(z)=\frac{\alpha}{2}|\phi(z)|^2-\frac{\beta}{4}|\phi(z)|^4-\frac{\gamma}{5}|\phi(z)|^5-\frac{V_0}{2}\textrm{sech}^3(z)|\phi(z)|^2$. To analyze the stability of the kink it is essential to calculate the effective force $F(z)$ where $F(z)=-dU(z)/dz$. One can now determine the stability of the kink by analyzing the zeros of $F(z)$ which can be noted from the Fig.\ref{force}. Let $z_c$ corresponds to one such point, i.e., $F(z_c)=0$. If $F(z)<0$ for $z<z_c$ and $F(z)>0$ for $z>z_c$ then $z=z_c$ is a stable equilibrium position for the kink. Here, we find the zeroes of $F(z)$ lies at $z_1=-0.47$ and $z_2=0.48$. Further, we see that, $F(z)<0$ for $z<z_1$ and $F(z)>0$ for $z>z_1$. Thus $z=z_1$ is a stable equilibrium position for the kink, whereas at $z>z_2$, $F(z)<0$ and $F(z)>0$ for $z<z_2$. Hence, we can conclude that the kink is in unstable equilibrium at $z=z_2$. In other words, the point $z_c$ where $F(z)$ changes the sign correspond to the equilibrium positions with $\frac{dF(z)}{dz}|_{z=z_c}>0$ being the stability criterion. Hence, the stability condition for the anti-kink can be noted as $\frac{dF(z)}{dz}|_{z=z_c}<0$ \cite{gonz}.

\section{Traveling Solution}\label{dyn}
In this paper, we have mainly focused on static solitonic solution for CQNLSE however in this section we like to explicate on the traveling solutions which are akin to the static counterpart. Starting from Eq.(\ref{eq1}), we consider $\psi(x,t)=\rho(x,t)e^{\theta(x,t)-i\mu t}$. However, we are interested in the real envelop solution at this point, hence the amplitude ($\rho(x,t)$)  and the phase ($\theta(x,t)$) functions are real. Inserting the ansatz in Eq.(\ref{eq1}) we can decompose it in $a+ib=0$ form so that we can separately investigate $a=0$ and $b=0$ equation which we name as real and imaginary equation respectively. The imaginary equation will read,
\begin{eqnarray}
-\mathcal{A}\left(2\frac{\partial\theta}{\partial x}\frac{\partial\rho}{\partial x}+\frac{\partial^2\theta}{\partial x^2}\rho\right)&=&\frac{\partial\rho}{\partial t}\nonumber\\
\frac{\partial}{\partial \xi}\left(-u\rho^2+\mathcal{A}\theta_{\xi}\rho^2\right)&=&0\nonumber\\
\frac{u}{\mathcal{A}}+\frac{C_0}{\mathcal{A}\rho^2}&=&\theta_{\xi}. \label{im}
\end{eqnarray}
Here, $\xi=x-ut$ describes the center of mass frame which is moving with a velocity $u$ with respect to the lab frame and $C_0$ is the integration constant. If phase and amplitude are not coupled then we can safely set $C_0=0$. Using Eq.(\ref{im}) we can now write the real equation such that,
\begin{eqnarray}
-\mathcal{A}\left(\frac{d^2\rho}{d\xi^2}-\theta_{\xi}^2\rho\right)+\mathcal{B}\rho^3(\xi)+\mathcal{C}\rho^4(\xi)+\mathcal{D(\xi)}\rho(\xi)&=&\mu\rho,\nonumber\\
\frac{d^2\rho(\xi)}{d\xi^2}+\tilde{\alpha}\rho(\xi)-\beta\rho^3(\xi)-\gamma\rho^4(\xi)-\delta(\xi)\rho(\xi)&=&0,\nonumber\\\label{re}
\end{eqnarray}
where $\tilde{\alpha}=\mu-u^2/\mathcal{A}$. Here, we have also translated the external potential to the center of mass frame from the lab frame following the recipe suggested in Ref.\cite{das3}. 
If we define the dimensionless length scale as $\eta=\zeta\xi$, where $\zeta^{-1}$ is the coherence length, then Eq.(\ref{re}) can be rewritten as,
\begin{eqnarray}
\zeta^2\frac{d^2\rho(\eta)}{d\eta^2}+\tilde{\alpha}\rho(\eta)-\beta\rho^3(\eta)-\gamma\rho^4(\eta)-\delta(\eta)\rho(\eta)&=&0.\nonumber\\\label{sre}
\end{eqnarray}
Now, it must be noted that Eq.(\ref{sre}) is very much similar to Eq.(\ref{eq3}). Thus we can safely use the functional forms of the solutions obtained in the previous section except they are now in the co-moving frame. 
\section{Conclusion}\label{con}
Motivated by the recent application of CQNLSE in ultra-cold atomic gases to describe the liquid-like phase observed in the experiments over the last couple of years, we analyze the possibility of obtaining cnoidal solutions in CQNLSE under the influence of an external trap. To obtain the cnoidal solutions, it is necessary to choose the external potential as a nonlinear cnoidal function. Our analysis reveals that, with cubic and quartic interaction it is unlikely that we can obtain any sinusoidal mode however localized (sech) and delocalized (tanh) modes can be obtained. A bright soliton-like solution is obtained under the influence of a  P\"oschl-Teller like potential (not exact P\"oschl-Teller potential). Though we obtained a unique $w$-soliton, however it turns out that physically, the existence of this soliton is unlikely. 
An asymmetric step-like potential results delocalized solitons which are popularly known as kink and anti-kink. Further, we discuss the stability of these localized and delocalized solitons. Later we describe the prescription to extend our investigation, which is static in nature, to obtain dynamical solitary waves. 

We must acknowledge that a detailed description regarding the droplet phase is of high interest, but it is beyond the purview of the current study. We will definitely come back to discuss the droplet formation in quasi one dimensional ultra-cold atomic gases very soon where our present understanding about the CQNLSE will play a major role. Considering the nascent stage of quantum liquid research we also expect our findings will be of high significance in analyzing this new phase of matter for the scientific community. This will enable us to characterize the droplets more precisely.

\section*{Acknowledgement} AK, thanks Science and Engineering Research Board (SERB) under Department of Science and Technology (DST), India for the support provided
through the project number CRG/2019/000108.
\section*{Author Contribution Statement} AK conceptualized the problem. AD carried out the basic calculations and afterwards both contributed equally in finalizing the results and writing the manuscript. 

\bibliographystyle{epj}
\bibliography{ms-v2}

\end{document}